\documentstyle[12pt,aaspp4]{article}

\def\approx{$\sim$}
\def\h2{$\rm H_2$}
\def\ammonia{$\rm NH_3$}
\def\nh3{$\rm NH_3$}
\def\water{H$_2$O}
\def\e#1{$\times 10^{#1}$}
\def\tenup#1{10$^{#1}$}
\def\asec{\arcsec}
\def\amin{\arcmin}
\def\deg{\arcdeg}
\def\thr{$^h$}
\def\tmin{$^m$}
\def\tsec{$^s$}
\def\kms{km~s$^{-1}$}
\def\solum{$\rm L_{\sun}$}
\def\solmass{$\rm M_{\sun}$}
\def\percc{$\rm cm^{-3}$}
\def\peryr{$\rm yr^{-1}$}
\def\percm{$\rm cm^{-1}$}
\def\ch2{$\chi^2$}

\begin{document}

\title{Radiative Transfer Modelling of the Accretion 
Flow onto a Star-Forming Core in W51}
\author{Lisa M. Young\altaffilmark{1}}
\altaffiltext{1}{Tombaugh Scholar}
\affil{Astronomy Dept., New Mexico State University, Box 30001, Las Cruces,
NM 88003; lyoung@nmsu.edu}
\author{Eric Keto and Paul T. P. Ho}
\affil{Center for Astrophysics, 60 Garden St., Cambridge, MA 02138;
keto@cfa.harvard.edu, ho@cfa.harvard.edu}

\begin{abstract}
We present an analysis of the temperature, density, and 
velocity 
of the molecular gas in the star-forming core around W51 e2.  
A previous paper (Ho \& Young 1996) describes the 
kinematic evidence which implies that the core around e2 is contracting 
onto a young massive star. 
The current paper presents a technique for modelling 
the three-dimensional structure of the core by simulating 
spectral line images of the source and comparing those images to observed 
data.  
The primary conclusions of this work are 
that the molecular gas in e2 is radially contracting at about 5 \kms\ and 
that the temperature and density of the gas decrease outward over 0.15~pc
scales.
The simple model of the collapse of the singular isothermal sphere 
for low-mass star formation (Shu 1977) is an inadequate
description of this high-mass molecular core; better models have temperature
$\propto r^{-0.6}$, density $\propto r^{-2}$, and velocity $\propto
r^{+0.1}$.
The core appears to be spherical rather than disk-like at the scale of
these observations, 0.3 pc.
In this paper we show how a
series of models of gradually increasing complexity 
can be used to investigate the sensitivity of the model to its parameters.
Major sources of uncertainty for this method and this dataset are the
interdependence of temperature and density, the assumed \ammonia\ abundance,
the distance uncertainty, and flux calibration of the data.
\end{abstract}


\section{Introduction}

In recent years, radio interferometric telescopes have provided a wealth of
data on the internal structure and dynamics of molecular clouds and star forming
regions on scales of 1 pc and smaller.  
For example, observations of neutral
gas have shown the presence of infall and spin-up motions, 
rotating disks, outflows, and expanding molecular shells at later stages in
the formation of stars (Ho \& Haschick 1986; Keto, Ho, \& Haschick 1987, 
1988; Torrelles et al. 1989;
Sargent \& Beckwith 1991; Carral \& Welch 1992; Torrelles et al. 1993;
Kawabe et al. 1993; and Ho \& Young 1996, hereafter Paper~I). 
The details of the collapse process are vital to understanding how stars
can be formed-- how angular momentum and magnetic flux are 
transported out and how
simultaneous accretion and outflow determine the mass of the resultant
star.

Theoretical models have mostly focused on the
formation of low-mass stars.
For example, Shu (1977) analyzed the
gravitational collapse of non-rotating, non-magnetic, isothermal gas 
spheres. 
He proposed an ``inside-out'' collapse in which 
the contracting portions of the cloud should develop an $r^{-1.5}$ density 
profile and an $r^{-0.5}$ radial infall velocity.  
More recently, Mouschovias and others 
considered the effects of magnetic fields; they found that under certain
common conditions the density will vary as 
$r^{-1.5}$ to $r^{-2}$ over scales of \tenup{-3} pc to a few tenths of a
parsec (e.g. Basu \& Mouschovias 1994). 
Scoville \& Kwan (1976)  studied the temperature distribution in a
centrally condensed cloud heated by a source of radiation such as an HII
region.
Assuming thermal equilibrium between the radiation and the dust, and
between the dust and the gas, they calculate that the
temperature of the gas and dust
should decrease with radius as $r^{-0.3}$ or $r^{-0.4}$.  
They expect to find temperatures of 50--70~K at
a distance of 0.07 pc from an object of luminosity \tenup5\ \solum, given
typical gas and dust conditions.
The present paper is part of a study to observationally measure some basic
properties of high-mass star formation and to determine whether the
results mentioned above, some of which are intended for low-mass stars, also
describe high-mass star formation.

Paper I presents observations of the \ammonia\ (J,K) = (2,2) 24 GHz inversion
transitions in the star forming region W51.  Those observations detected an
accretion flow of a few \kms, extending over about 0.3 pc in diameter, 
onto a young star.
The star must be massive because it has created an ultracompact HII
region, called W51 e2, which is embedded in the molecular core.
The current paper takes a numerical modelling approach to finding the
structure of the star forming core in W51.  
Paper~I's observations of W51 e2 are compared with theoretical spectra
that would be observed from a model with a specified temperature, density,
and velocity structure.
We begin with the simplest model with the fewest parameters,
that of a uniform density isothermal sphere, and
show how the addition of an infall velocity and temperature and
density gradients improves the fit to the data. 
This approach constrains the physical conditions of the molecular gas in
the core; it also has the important advantage of revealing how tightly the
present data and models can constrain those conditions.

\section{Background} 

\begin{figure}
\plotfiddle{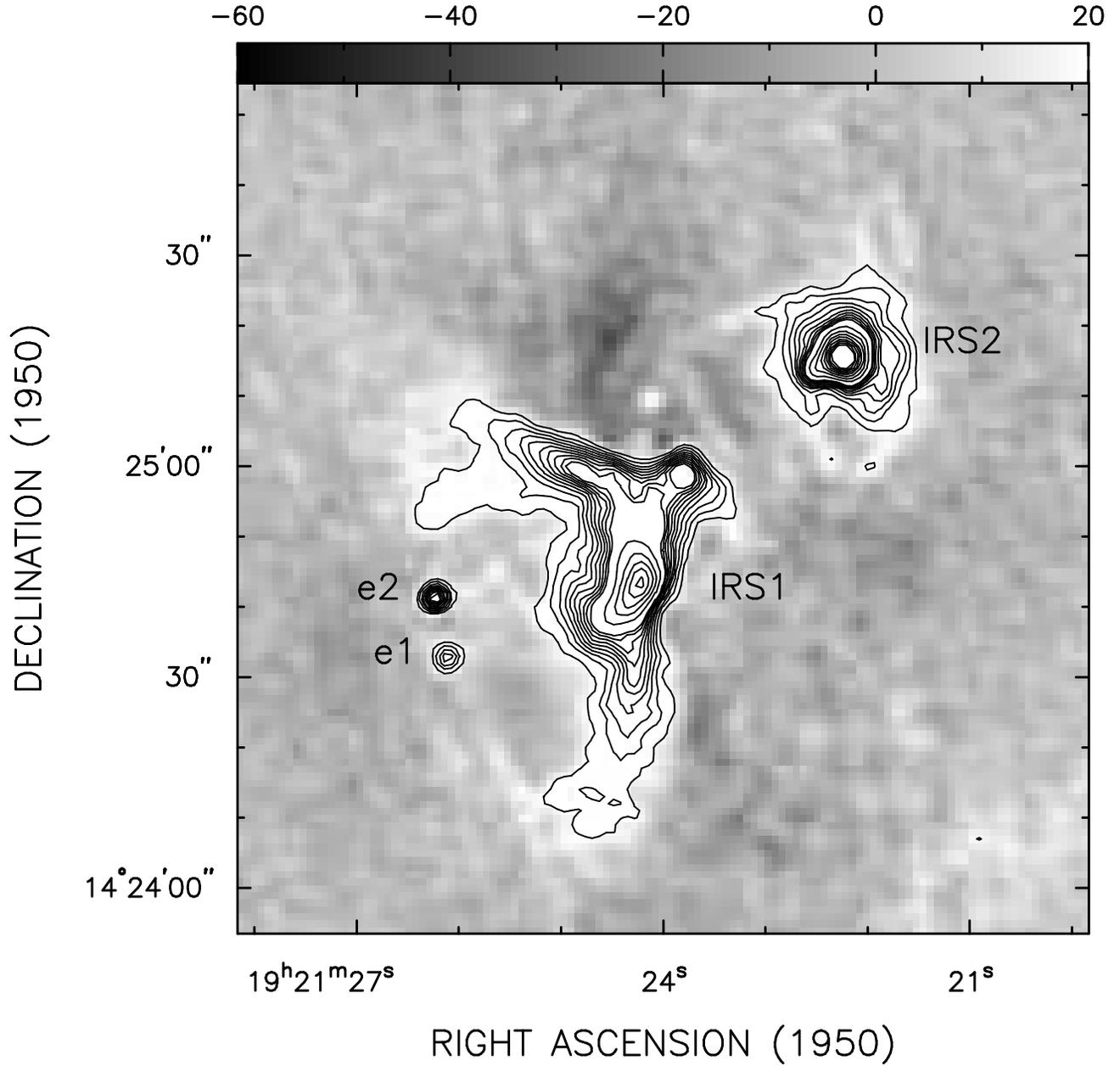}{7 in}{-90}{100}{100}{-300}{600}
\caption{
Continuum emission at 1.3 cm from W51,
taken from Paper 1.
Contour levels are (1, 2, 3, 4, 5, 6, 7, 8, 9, 10, 15, 20, 25, 30, 35, 40,
45, 50, 100) $\times$ 20~mJy/beam or 6.4~K.
The greyscale image is also the continuum emission from 20~mJy/beam to
$-$60~mJy/beam as shown in the scale at top.
The resolution is 2.6\asec.
\label{continuum}}
\end{figure}

W51 is an active region of high mass star formation, as shown by its
large luminosity (3\e{6} \solum, Thronson \& Harper 1979), IR objects
(Genzel et al. 1982; Bally et al. 1987), \water\ masers (Schneps et al.
1981), and shocked \h2\ emission (Beckwith \& Zuckerman 1982).  
Genzel 
et al. (1981) used the method of statistical parallax of \water\ masers around the
HII region W51 e2 (Scott 1978) to determine a 
distance of 7.0$\pm$1.5 kpc, and similar distances have been found for the
other objects in the W51 complex.
This paper concerns the ultracompact HII region
W51 e2 and the molecular core surrounding it.
Figure~\ref{continuum} shows the 1.3~cm continuum emission in 
W51.
Scott (1978) noted that
the flux from the HII region in e2 could be accounted for by the presence
of one ZAMS star of spectral type B0--O9.  
HII regions e1 and e2 also are surrounded by condensations of warm \ammonia, detected in both
emission and absorption by Ho,
Genzel, \& Das (1983).  
Though outflows are seen in other parts of the W51 complex, there is little
evidence of outflow activity near the e2 core on the arcsecond
scales of interest here (Mehringer 1994; Zhang \& Ho 1997).

\subsection{Data}

The observations for Paper I and the data reduction techniques used
are described in detail in that paper.  
Briefly, we used the NRAO Very Large Array (VLA)\footnote{The National
Radio Astronomy Observatory is a facility of the National Science
Foundation operated under cooperative agreement by Associated Universities,
Inc.} 
to observe the \ammonia\ (J,K) = (1,1) and (2,2) inversion transitions at
23.69 GHz and 23.72 GHz.  
The bandwidth of 6.25 MHz covers the
main quadrupole hyperfine component and all four satellite components with
64 velocity channels.  
The velocity resolution of this data is 1.24 \kms, and the bandpass was centered on a
velocity of +60 \kms\ with respect to the local standard of rest (LSR).
The images studied in this paper were made with a
spatial resolution (FWHM) of 2.6\asec, or 0.09
pc at a distance of 7.0 kpc.  
The rms noise level in the line-free channels
is 6 mJy/beam = 1.9~K. 

\subsection{Prediction of geometry in the core e2} \label{data}

Figure~\ref{comparetothis} presents a position-velocity diagram made along
a slice through the e2 molecular core;
the models of this paper attempt to reproduce the features shown in this
diagram.
(Paper~I presents many additional figures showing the distribution and
velocity structure of the molecular gas in and around W51 e2.)
A good model of the molecular gas in the e2 core should reproduce the
following features which are visible in Figure~\ref{comparetothis} and in the
figures of Paper I.

1.  The five hyperfine components of the transition are visible in both
emission and absorption because the absorption is consistently redshifted
with respect to the emission.
The absorption corresponds to cool (relative to the HII region) molecular gas in front 
of the HII region, whereas 
molecular gas to the side or in back of the HII region
is in emission.  

2.  The central hyperfine component of the emission line in e2 shows a curvy 
``C'' shape.
Emission east and west of the HII region (upper and lower edges of the
panel; see also Paper I) is at a velocity of about 55 \kms.
Towards
the center of the panel (the center of the HII region and molecular core) 
the emission line becomes more blueshifted. 
The line center and systemic velocity of the core seem to be at 54--55
\kms, which is about halfway between the emission
and absorption, at 50 and 60
\kms\ respectively.

3.  The ``C'' shape appears in any position-velocity diagram that is made
through the core e2, regardless of orientation.
Thus, position-velocity diagrams through e2 show approximate
circular symmetry on the sky.

4.  The position-velocity diagrams that just
miss the HII region  (see Paper~I) do not show absorption or a
curved C-shaped emission line, but they
do show that the lines are wider at the position of the HII region than,
for example, east or west of it.
In other words, this feature is an increase in line width at small spatial
scales.

5.  Optical depths in the (2,2) transition in e2 are high;
the ratio of the central hyperfine emission
component to the satellites is 2.5:1 to 3:1 in the core (Paper I),
implying emission optical depths of 7--10 (Ho \& Townes 1983).
In absorption all five hyperfine components have approximately the same
strength, implying very high optical depth.

6.  In e2 the peak of emission and the peak of absorption are not spatially
coincident, as might be expected for a perfectly spherically symmetric core.
Instead the emission and absorption peaks are offset by 3\asec\ or 0.1 pc,
suggesting that the HII region is off-center with respect to the molecular
gas or that the properties of the gas are not spherically symmetric.
This offset is confirmed by higher resolution observations (Zhang \& Ho
1997).

Paper I proposes a simple explanation, summarized here, that explains
these observed features.  The molecular core e2,
about 0.13 pc in radius, is a roughly spherical cloud of gas which is 
contracting onto a young massive star and HII region near the 
center of the sphere.  Thus the front of the cloud is moving away
from us and the back is moving toward us, as required by the redshifted
absorption and blueshifted emission; the ``C'' shape is a projection effect.
The assumption of a roughly spherical contraction explains the approximate
circular symmetry in the plane of the sky (point 3 above).
Paper I discusses this interpretation in more detail.

\section{Procedure} \label{procedure}

\begin{figure}
\plotfiddle{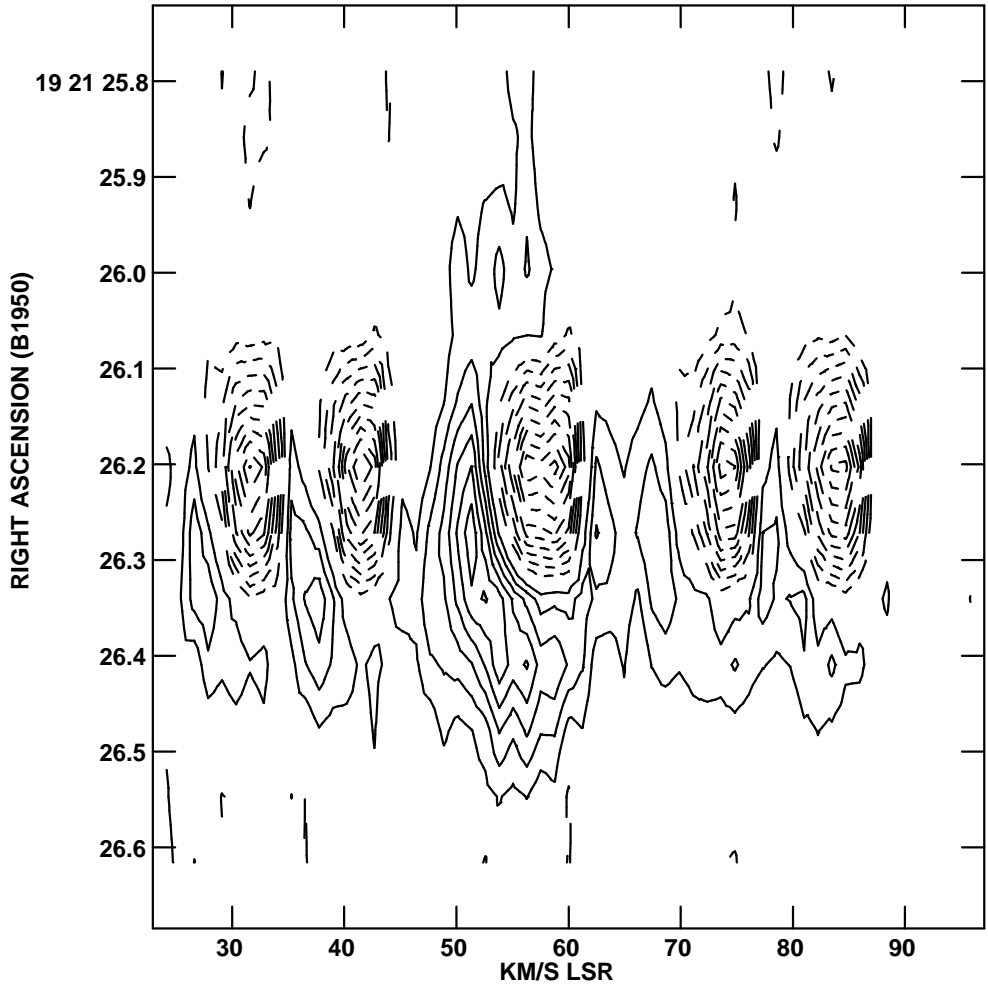}{5 in}{0}{70}{70}{-250}{-100}
\caption{Position-velocity diagram showing the \ammonia\ (2,2) data 
that are modelled in this paper.
Contour levels are (-10, -9, -8, -7, -6, -5, -4, -3, -2, -1, 1, 2,
3, 4, 5, 6, 7, 8, 9, 10) $\times$  0.012 Jy/beam.  
This velocity-right ascension slice was made at a declination of 14\deg\
24\amin\ 41\asec.
(Paper~I shows additional position-velocity diagrams through the e2
molecular core.)
Before modelling, the data in this image were subsampled and pixels showing 
mostly noise were trimmed off around the edges.
The result of the trimming and subsampling is shown in the top left panel
of Figure~\ref{bigfigpart1}.
\label{comparetothis}}
\end{figure}

\begin{figure}
\vskip 0 in
\plotfiddle{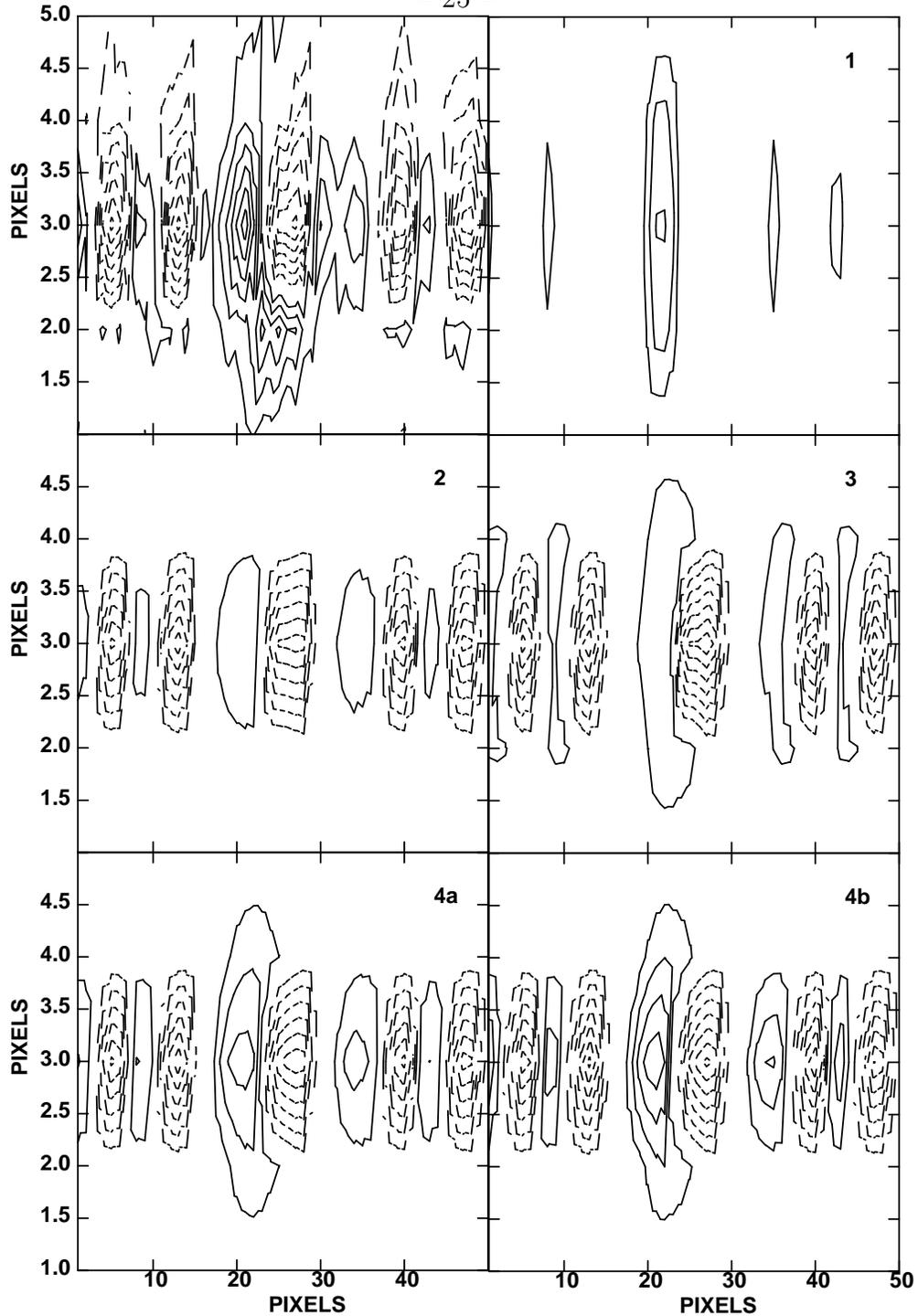}{6.8 in}{0}{77}{77}{-250}{-53}
\caption{
Data and models.
The top left plot shows the data with which the models are compared.
This observed position-velocity diagram is the same data as
Figure~\ref{comparetothis} but it has been subsampled so that the spatial
pixels are one beam width apart, hence independent of each other.
Contour levels for all panels are the same as in Figure~\ref{comparetothis}.
The axis on the left side is marked in
pixels of increasing RA, where one pixel is now 2.6\asec\ (0.09 pc) across.  
The axis
along the bottom shows pixels of velocity; high velocities (redshifted gas)
are on the right-hand side of the plot.  
The other plots in the figure show optimized output of the model number
indicated in each top right corner.
\label{bigfigpart1}}
\end{figure}

We investigate the structure of the molecular core around e2 by radiative
transfer modelling of the \ammonia\ emission.
Because the data show approximate circular symmetry in the
plane of the sky (Section~\ref{data}), 
we modelled only one two-dimensional slice, or position-velocity diagram,
through the e2 core.
Figure~\ref{comparetothis} is the position-velocity diagram
selected for modelling; it is the (J,K) = (2,2) transition,
and passes through the center of the HII region.
Figure~\ref{comparetothis} shows many of the features described in Section~\ref{data}.
The data in Figure~\ref{comparetothis} were spatially subsampled by
taking five pixels separated by
the FWHM of the beam, and were trimmed in velocity by  
selecting the central 50 of the observed 64 velocity channels.
The result of the subsampling is shown in the top left panel of Figure~\ref{bigfigpart1};
it is made up of 250 independent data points.

\begin{figure}
\plotfiddle{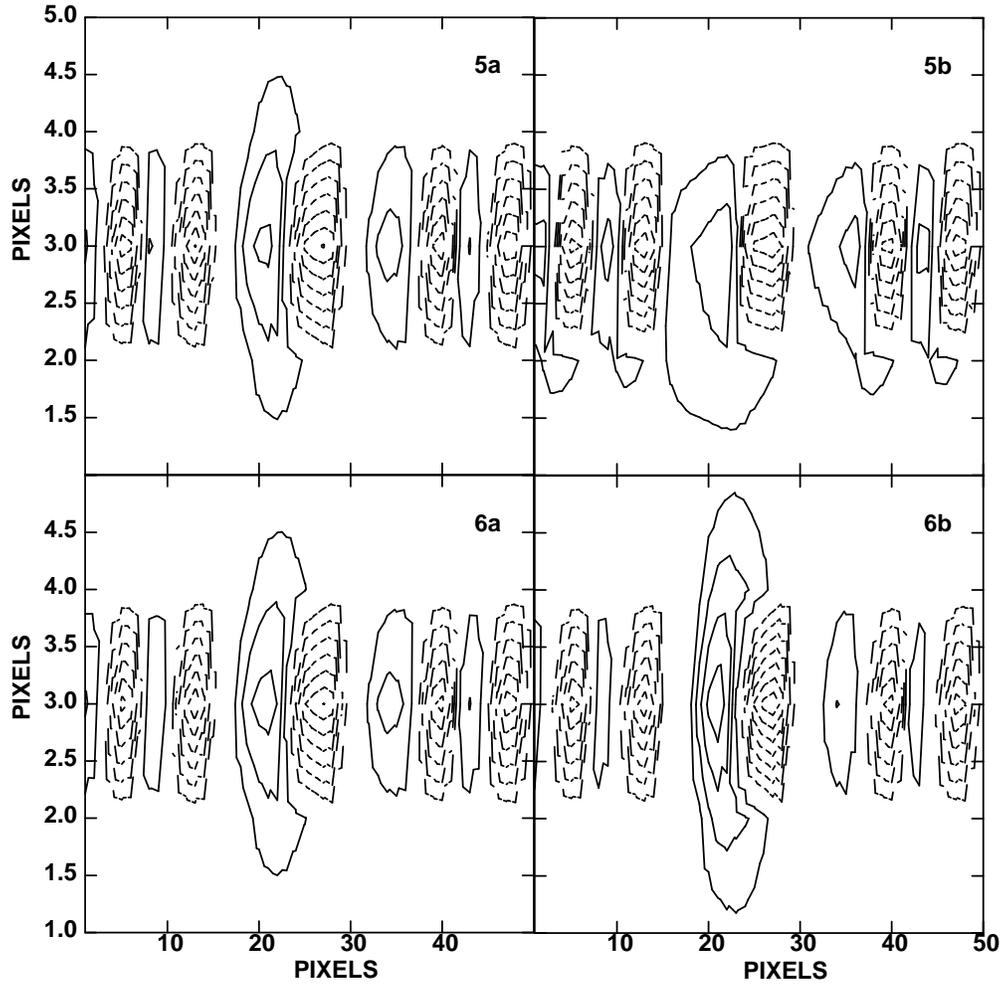}{4 in}{0}{77}{77}{-250}{-150}
\caption{
Same as Figure~\ref{bigfigpart1}, but for models 5 and 6.
\label{bigfigpart2}}
\end{figure}

The radiative transfer code used in our spectral line modelling is described
briefly in Keto (1990).  
Based on Paper~I's discussion of physical conditions in e2, we determined
an initial guess of the structure--- temperature, density, and velocity
field--- of the core.
For simplicity,
the physical parameters are described as power-law parametrizations in radius.
Level populations of \ammonia\ are determined using the assumption of local thermodynamic
equilibrium (LTE), and the line brightness is computed by integrating the
radiative transfer equation along the line of sight.
The calculated line radiation is convolved to the resolution of the
observed data and converted to the same
physical units as the observed data.
Thus the models in Figures~\ref{bigfigpart1} and \ref{bigfigpart2}
are sampled at the same spatial frequency as the data in the top left panel of 
Figure~\ref{bigfigpart1}, and their intensities are directly comparable.

In addition, we have added for this project a
least squares fitting procedure to optimize the modelled physical
conditions.
The multidimensional least-squares fit is done using a downhill 
simplex algorithm (Press et al. 1993).  
This algorithm is a gradient descent procedure which reaches a local
minimum but not necessarily a global minimum.
The fitting routine imposes no constraints on ``reasonable'' or
``acceptable'' physical conditions or power-law slopes aside from the
requirements that gas temperatures exceed 3~K and densities exceed 10
\percc.
Energetic and dynamic consistency of the models are considered in
Section~\ref{realitychecks}.

The radiative transfer/model fitting code also performs a
simple error analysis.
It estimates the error in each parameter as the second derivative of \ch2\ 
with respect to the parameter 
using the values of \ch2\ at the optimized value of the parameter and at two
symmetrically offset values of the parameter (Bevington, 1969).
This procedure for estimating the error assumes that the parameters are uncorrelated and that
the model is linearly dependent on them.
Neither of these conditions are true; however, comparison with a limited
Monte Carlo analysis indicates that our derived errors are at least of the
correct magnitude. 

Subsequent sections present the results of 
the data fitting for the molecular core e2, 
employing a series of
models of gradually increasing complexity. 
For example, the first model is a spherical cloud of molecular gas of constant
temperature and density and an HII region inside.  
In successive models, parameters allowing for
infall velocity and for gradients in the temperature, density, and velocity
are added.
There is no evidence for rotation in e2 on the scales of interest
here (Paper I), so the models do not include rotation.
(Zhang \& Ho [1997] found evidence of spin-up in e2 but only at radii
less than 0.2\asec, much smaller than the 1.3\asec\ resolution used for the
current study.)
The technique of gradually increasing the complexity of the models proves
extremely valuable because comparisons between the models reveal (1) which
parameters are important and which are not; (2) how well determined are the
physical conditions in the core. 

\section{Results}

\subsection{Model 1: quiescent cloud}
\label{model1section}

The first model consists of an 
HII region surrounded by a spherical cloud of molecular gas with 
uniform density and temperature, and no infall velocity.
A turbulent line width (FWHM) of 1.25 \kms\ in the molecular gas is assumed, 
based on the
observed line width in an optically thin envelope of gas surrounding e2 and
e1 (Paper I).
We also assume a fractional abundance \ammonia/\h2 = 1.4\e{-6}
in order to translate from the \ammonia\ density, which is constrained by the 
data, to the \h2\ densities quoted in this paper.
This \ammonia\ abundance is based on modelling of a similar 
high-mass star formation region, G10.6--0.4 (Keto, Ho, \& Haschick 1988). 
Abundances around \tenup{-6} are also estimated for the \ammonia\ near
G9.62+0.19 and G29.96--0.02 (Cesaroni et al. 1994).
The turbulent line width and \ammonia\ abundance remain fixed for all models.
Model 1 has six free parameters:  the systemic velocity
of the HII\ region,
the continuum opacity of the HII region, 
the radius of the HII region (taken to be the inner radius of the
molecular gas),
the outer radius of the molecular cloud, and the temperature and density of 
the gas in the cloud.  
This model provides a
null hypothesis for comparison to models with contracting molecular gas.

Table~\ref{parameters} gives the initial guesses of the parameters
describing the gas in this and subsequent models;
Table~\ref{results} gives the optimized parameters and reduced $\chi^2$
(per degree of freedom) for all models.
The initial guesses are presented because the more
complicated models sometimes give substantially different output from
different sets of initial conditions; this issue is discussed more fully in
Section~\ref{model4section}.
Table~\ref{maxmin} also gives maximum and minimum values for the gas
temperature, density, and velocity.

Figure~\ref{bigfigpart1} presents the best fit 
position-velocity diagram for model 1 and subsequent models.  
The panel labelled ``1" should be compared to the data in the top left
panel of the same figure.
The most obvious problem with this model is that because infall (or
outflow) is not allowed, emission and absorption
are constrained to have the same line width and same radial velocity.
In this fit, the radius and continuum opacity of the HII region are 
consistent with zero: 0.003$\pm$0.06 pc and (0.02$\pm$3)\e{-19} \percm.
The fitted optical depth of the HII region is only 3\e{-5}, so low that 
absorption is not seen.
The value of the reduced \ch2\ (per degree of freedom) for this model is
11.2, which is little better than the \ch2\ of blank sky 
(12.2).

\subsection{Model 2: contracting cloud with uniform temperature and density}
\label{model2section}

The second model incorporates all the features of the
first model, namely an HII region surrounded by a spherical cloud of
gas with uniform temperature and density, and adds a radial infall 
velocity of the form
$v = v_0 (r/{r_0})^{\alpha_v}$.
Model 2 fits eight free
parameters:  $v_0$, $\alpha_v$, and the six parameters of model
1.

Again, Figure~\ref{bigfigpart1} and Table~\ref{results} present the results 
of this optimization.
Clearly, the addition of infall velocity improves the model immensely.
The value of reduced \ch2\ drops by almost a factor of three, to 3.8.
The fitted continuum opacity of model 2 is 1.4\e{-19}~\percm,
producing a continuum optical depth of 0.02 for the HII region.
(Subsequent models have very similar continuum optical depths. 
However,
we caution that these continuum parameters are poorly constrained because
the HII region is not resolved by these observations; see Gaume, Johnston, \&
Wilson 1993.)
In contrast to model 1, the gas in front of the HII region is now
redshifted and is seen in absorption. 
The pattern of redshifted absorption and blueshifted
emission, so obvious in the data, is now reproduced by the model as well.

\subsection{Model 3: Shu (1977)}
	
The simple analytic solution of Shu (1977) for the properties of a
collapsing molecular core can be directly tested using our observations of
W51 e2.
Since that analytic solution was developed specifically for the case of low
mass star formation, the relevance of the solution for the e2 core is not obvious.
However, the Shu (1977) solution is included as model 3 because comparisons
between model 3 and models 2, 4a, and 4b help disentangle the importance of the
various parameters.
Model 3 is similar to model 2 except for the addition of a radial 
gradient in density, fixed as $n \propto r^{-1.5}$. 
In addition, the infall velocity is fixed at $v \propto r^{-0.5}$, and 
there is no gradient in temperature.
Thus, model 3 has seven free parameters: the eight of model 2, minus the
slope describing the gradient in infall velocity.

The minimized value of reduced \ch2\ for model 3
is greater than that for model 2 by 0.3. 
As for model 2, also an isothermal model, the emission main/satellite 
hyperfine intensity ratio is too
low and the emission is not strong enough, which would be rectified by 
the addition of some hotter and optically thinner molecular gas.
Thus Shu's low-mass stellar collapse model is not an adequate description 
of the collapse of the high-mass e2 molecular core.
An obvious reason is the increased importance of central heating in the
high-mass case.
Subsequent models return to fitting the density, velocity, and
temperature slopes as free parameters.

\subsection{Model 4: gradients in temperature and density}
\label{model4section}

Model 4 incorporates the features of model 2 and adds radial gradients in
temperature and density.
Radial gradients are expected to improve the fit to the data for the
following empirical reason.
The data (Figure~\ref{comparetothis}) show stronger emission in 
the main hyperfine component than in the satellites, whereas models 2 and 3 
produce about the same intensity in all emission components.
Density and temperature gradients allow the introduction of some hotter,
optically thinner gas, which would increase the relative strength of
the central hyperfine component.
Model 4 has 10 free parameters: the same eight as model 2,
plus the exponents in the temperature and density power laws.

Table~\ref{results} presents the results of two optimizations of model
4, and the corresponding position-velocity diagrams are
both presented in Figure~\ref{bigfigpart1}.
The difference between model 4a and model 4b is simply the initial guess
(Table~\ref{parameters});
model 4a starts with a higher density and a lower temperature than model 4b.
The reason for running these two cases is that we know the brightness of a
single molecular line cannot be used to uniquely determine both the gas 
temperature and density.  
In the optically thin case, for example, the line brightness temperature is
the product of the temperature and optical depth.
Thus, the two different models 4a and 4b allow us to gauge how well
temperature and density can be constrained.
Both models 4a and 4b are
significant improvements over the models 2 and 3; their values of
reduced \ch2\ are 3.2 and 3.0, respectively, compared to 3.8 for model 2
and 4.1 for model 3.
The difference in \ch2\ between 3.2 and 3.0 is not significant.

The optimized infall velocities are not very different in models 4a and 4b
from the infall velocities in model 2 (Table~\ref{maxmin}), which implies 
that the infall velocities are well constrained in this technique.
Comparing models 4a and 4b to models 2 and 3,
the central emission components are stronger and
the main/satellite intensity ratios are higher.
The emission also has a larger spatial extent
in models 4a and 4b than in model 2.
In the optimized models 4a and 4b,
molecular gas densities drop by two orders of magnitude between the
outer radius of the HII region and the outer radius of the cloud; the
temperatures drop by about 10--15 degrees.
Thus, a good description of
the core e2 requires radial gradients (decreasing outwards) in temperature 
and/or density.
Model 3, which is isothermal but has a density gradient, suggests that a 
radial gradient in density alone is not sufficient for a model of e2; a
gradient in temperature is also required.
Of course, a temperature gradient should not be
surprising since there is a heat source (the star) in the center of the core.
Analysis of the low-mass star-forming core B335 (Zhou et al. 1993) also
shows evidence for a temperature gradient in that core.

As expected, the results of models 4a and 4b show that the 
temperature and density are not independent parameters; to some extent, a
lower temperature can be compensated by a higher density.
Fitting two \ammonia\ transitions simultaneously, such as (1,1) and
(2,2), would remove this ambiguity.
Uncertainties are discussed further in Section~\ref{uncertainties}.

\subsection{Models 5a and 5b:  offset HII region} \label{model5section}

The modest asymmetries in the observed \ammonia\ emission suggested that
a better fit to the data might be achieved by displacing the HII region 
a few arcseconds (up to 0.1 pc) from the center of the spherical molecular
core (Section~\ref{data}).
Models 5a and 5b elaborate on models 4a and 4b by including the 
position of the HII region within the core as
free parameters. 
The molecular gas temperature is calculated with respect
to distance from the HII region, the heat source. 
Density and infall velocity are calculated with respect to distance
from the center of the spherical core, as the young star's mass is much
less than the mass of molecular gas in the core (Paper I; Zhang \& Ho 1997).
This rather simplistic model has the advantage that it introduces some 
asymmetry without adding many new free parameters.
Models 5a and 5b have twelve free parameters:  the same ten from models 4a
and 4b, with the addition of the HII region's offset in two
directions (along the line of sight and the direction of right
ascension).

As for models 4a and 4b, models 5a and 5b differ in their initial guess
parameters.
The results of these optimizations are given in Figure~\ref{bigfigpart2} and
in Table~\ref{results}.
For model 5a, the initial guess offset of the HII region is 0 pc, and for
model 5b the initial offsets are 
0.05 pc in each of the two directions, 
chosen to agree with the asymmetries in the data.  
Neither the optimized model 5a nor 5b achieves a significant improvement in
reduced \ch2\ over models 4a and 4b.
Model 5b better matches the east-west asymmetry of the data.
However, in both models 5a and 5b the optimized offsets are not significantly 
different from the initial guesses. 
It is possibile that our downhill simplex procedure failed to
optimize this particular model, despite its success with the others.
More likely, this result suggests that the model of the HII region offset 
from the center of its parent accreting cloud is not correct in the case of e2.
The asymmetry might be better described by a different model.
For example, there could be an overall east-west density gradient in the
molecular core around e2.  
Another possibility is that the molecular core
might not be spherical; we explore this possibility in models 6a and 6b.

\subsection{Models 6a and 6b:  molecular disk}

Because a non-spherical cloud model might 
help reproduce some of the east-west asymmetry in the observations of e2,
models 6a and 6b describe the molecular gas in e2 as a disk rather than a
sphere.
The disk is simply an oblate spheroid whose 
unique axis is confined to lie somewhere between 
the line of sight and the right ascension axis (the direction of 
the position-velocity diagram).  
Since only one position-velocity diagram is modelled, only one
inclination angle is required to specify a unique orientation of the disk.
Models 6a and 6b have 12 free parameters: the same 10 from  
models 4a and 4b, with the addition of the axial ratio of the 
spheroid and the inclination angle.  
The position of the HII region is fixed at the center of the disk.
No constraints are placed on the axial ratio or inclination of the disk,
but the approximate observed circular symmetry 
(Section~\ref{data}) implies that a highly inclined thin disk 
is an unreasonable solution.

Again, the two disk models 6a and 6b differ in their initial guess
parameters. 
Model 6a had an initial aspect ratio of 1:1, and its optimized parameters
are quite similar to those of models 4a and 5a.
Model 6b started with an aspect ratio of 4:1 and an inclination of 45\deg\
(0\deg\ is face-on);
its result is a flat disk (10:1), with an inclination of 0\deg.
Figure~\ref{bigfigpart2} shows that model 6b does the best job of all
models in reproducing the large spatial extent of the central emission
component.
Model 6b also has the highest central-to-satellite intensity ratio in
emission, which probably results from the relatively high temperatures
in this model (Table~\ref{maxmin}; Section~\ref{notreproduced}).
However, neither model 6a nor 6b has significantly lower \ch2\ than the simpler
models 4a and 4b or the offset HII region models 5a and 5b.
Furthermore, as in the case of the offset HII region, the two disk models do not
converge to the same result.
These facts suggest that the aspect ratio and inclination of
any putative disk structure are not well constrained by the current
procedure.

\begin{figure}
\plotfiddle{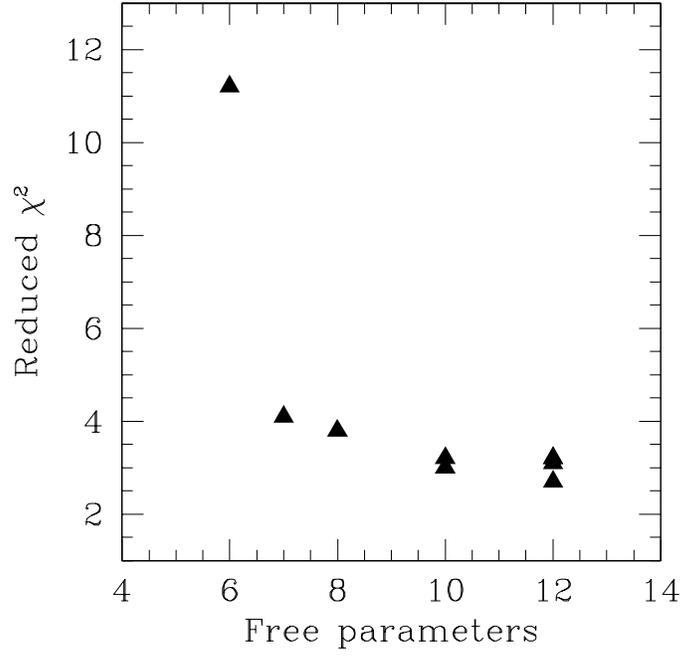}{3 in}{0}{100}{100}{-250}
{-100}
\caption{Reduced \ch2\ for each model compared with the number of free
parameters.  Improvements are realized until there are ten free
parameters (models 4a and 4b), but models 5a through 6b make no further
improvement in \ch2.
\label{ericplot}}
\end{figure}

On the philosophy that we should adopt the simplest model which best
describes the data, we plot reduced \ch2\ against the number of parameters
in each model (Figure~\ref{ericplot}).
This figure shows that as parameters are added, the fit of the models to
the data improves until model 4 is reached.
Models 5 and 6 increase the complexity of the model without significantly
improving the fit to the data.
Nevertheless, there are still a number of features of the data which are
not reproduced by the models (Section~\ref{notreproduced}).
We infer that the e2 core cannot be described as a simple 
sphere, but the specific asymmetries described by models 5 and 6 are not
required nor ruled out by the data.
We should therefore base our conclusions on model 4, which requires the
presence of infall and a centrally condensed and centrally heated molecular
core.

\subsection{Quantitative results} \label{quantitative}

Most of the optimized density exponents are close to $n \propto r^{-2}$.  
This slope is steeper than most theoretical predictions for low-mass
stars, which give
$n \propto r^{-1.5}$ within the region of contraction (e.g. Shu 1977).
However, a slope of $-$2 also agrees with the empirical results of Zhang \&
Ho (1997) for the W51 e2 core. 
They used higher resolution VLA observations to fit the column
density of \ammonia\ versus radius in e2 and find $n \propto r^{-2.0^+_-
0.1}$ within 5\asec\ (0.2 pc) of the HII region.
This agreement between the empirical results and radiative transfer modelling
gives additional confidence in the modelling technique.
Of course, the model results are based on the assumptions of LTE and constant
\ammonia\ abundance. 
Uncertainties introduced by these assumptions are discussed further in
Section~\ref{uncertainties}.

There are two models with density slopes quite different from $-$2.  
In model 4b the density falls off quite steeply
($r^{-3.9}$), which might be due to a trade-off between its high
temperatures (relative to the other models) and density.
In model 5b the density increases with larger radii ($r^{+0.9}$),
which is dynamically unstable and physically unrealistic.  
This unusual result probably
comes from calculating the density with respect to the center of the
sphere whereas the HII region is in fact offset by 0.07 pc from that center
in this model.

The radiative transfer models also have the  
temperature falling off a bit more steeply than expected.  
Models 4a, 5a, 5b, 6a, and 6b have
$T \propto r^{-0.6}$, whereas Scoville \& Kwan (1976) predicted
$T \propto r^{-0.4}$.  
In contrast, Zhang \& Ho (1997) find no evidence of temperature
gradients in the central 5\asec\ (0.2 pc) of the e2 molecular core.
Formal errors for the temperature exponent 
(see Table~\ref{results}) are consistently close to 0.05.
However, those formal
errors are most likely an underestimate of the true uncertainties
(Section~\ref{uncertainties}), so the temperature gradients in e2 may be
consistent with theoretical models.  
Since the isothermal models (numbers 2 and 3) have significantly higher \ch2\ than
those that fit a temperature gradient, we conclude that an isothermal model
is firmly ruled out by the radiative transfer fitting technique.
It is not clear why this modelling should give a different slope than 
Zhang \& Ho (1997) find, except perhaps for the fact that their result is based on
line ratios whereas the current technique uses essentially the beam-diluted
brightness temperature.

Those models which fit an infall velocity gradient 
have slopes between $v \propto r^{0.2}$ and $v \propto r^{0}$. 
Even models 2, 4a, and 6a, whose initial slope was $-0.5$, have optimized
slopes greater than zero.
Those slopes are not consistent with the ``inside-out" 
scenario proposed
by Shu (1977), in which the infall velocity must decrease with increasing
radius.
In a different star-forming core, however, an inside-out collapse has been
inferred.
In G10.6--0.4, the infall velocity decreases with radius at least as
quickly as $v \propto r^{-0.5}$ (Keto, Ho, \& Haschick 1988).
If an inside-out, accelerating collapse were present in W51~e2 as well,
we would expect to observe higher velocities on smaller spatial scales---
at least 10 \kms\ at the 0.01 pc scales observed by Zhang \& Ho (1997).
However, such high velocities are not observed.
This result could be related to the fact that the HII region is actually offset
from the center of the molecular core.

Table~\ref{maxmin} presents maximum and minimum values of the temperature,
density, and infall velocity in each model, as well as the total gas mass.
The extrema are calculated at the inner and outer edges of the shell of
molecular gas, as appropriate for each model.
From this table we see that the models all have infall velocities 
of 4--6 \kms, which are
consistent with the value inferred in Paper I. 
Such velocities are about a factor of 10 higher than the isothermal sound
speed (0.5~\kms\ for molecular hydrogen at 50~K).
Basu \& Mouschovias (1994, 1995) have theoretically analyzed the collapse of 
magnetized molecular cloud cores with ambipolar diffusion, and they predict 
infall velocities close to the sound speed, rather than an order of
magnitude higher than the sound speed.
The reason for this discrepancy is not clear, though Basu \& Mouschovias
(1995) state that less efficient coupling between neutrals and ions can
give rise to higher infall velocities.

Molecular hydrogen densities calculated for the radiative transfer models range
between $n$ \approx 5\e4 -- 5\e7 \percc\ (Table~\ref{maxmin}).
The critical density for exciting these \ammonia\ 
transitions is about \tenup4 \percc\ (Ho \& Townes 1983), so in this sense
the fitted densities are consistent with expectations.
There is a 
considerable range in the estimates of the density of
the gas, especially near the outer edges of the core (Table~\ref{maxmin}) where values
differ by three orders of magnitude.
The uncertainties in 
the density of the gas are large because, in the optically thick case,
small errors in
fitting the strength of the hyperfine components translate into large
errors in the density (see also Section~\ref{uncertainties}).  

All of the models have gas temperatures
which are lower than calculated from other techniques.
The (1,1) to (2,2) line ratios in the gas
surrounding e2 imply temperatures of 25--35~K at 0.2--0.3 pc from the
HII region e2 (Paper I). 
Zhang \& Ho (1997) also use high angular resolution line ratios to find 
temperatures of 40--50~K inside the core (inside 0.2 pc) and temperatures of 
25--30~K outside the core, at $\geq$ 0.2 pc from the HII region.
In contrast,
the models have peak temperatures of only 20--40~K and values of a 
few to 10~K at 0.2 pc from the HII region.
As discussed in Section~\ref{uncertainties}, factors of two in the
temperature are within the uncertainties caused by an inverse correlation
between temperature and density.
Furthermore, if the molecular gas is clumpy,
a beam filling factor less than one would make the modelled
temperatures, which are essentially derived from the brightness temperature
of the gas, lower than the excitation temperatures.
From the ratio of the observed continuum flux to the absorption line
strength, the beam filling factor for the redshifted absorbing gas 
is 0.8 (cf. Keto, Ho, \& Haschick 1987).

\section{Discussion}  

\subsection{Consistency checks} \label{realitychecks}

The only physical constraints placed on the model molecular cores are that the
temperature of the gas must be above 3~K and the density must be above
$n$ \approx 10 \percc.   
In other words, the models are fit without regard to 
energetic or dynamic self-consistency.  
Thus, some simple consistency checks are in order.
If the collapse begins from a
state in which the gas is stationary, cold, and essentially
infinitely far from the central star, 
the total of the
potential, kinetic, and thermal energy should be zero at every radius.

Table~\ref{energybal} gives the total energy in the  
molecular gas in the form of gravitational potential energy, infall kinetic
energy, and thermal energy.  
(The potential energy is the usual integral of
$-GM(r)m(r)/r$, where $M(r)$ is the total mass inside $r$ 
and $m(r)$ is the mass at $r$;
the kinetic energy of infall is 
the integral of $mv^{2}/2$, and the thermal energy is 
$3kT/2$ per molecule.)
The thermal energy in the gas is always dominated by the turbulent energy
corresponding to the assumed intrinsic linewidth of 1.25 \kms\
(540~K).
In turn this assumed turbulent energy is always less than the kinetic
energy of infall.
In most cases the sum of kinetic and thermal energy is within a factor of
2--3 of the potential energy, indicating approximate energy balance.
The exceptions to this statement are models 2 and 3, which are 
rejected in any case because of their relatively high values of \ch2,
and model 5b, which has the density
increasing outwards.

The total mass of gas in each model 
appears in Table~\ref{maxmin} and varies from 100 to \tenup4\ \solmass.  
These masses
are consistent with the 100 to 200 \solmass\ lower limit inferred in Paper
I, based on the assumption that the gas is moving at the free fall
velocity.
It is also possible to estimate a mass infall rate from the molecular
core using the density, velocity, and radius values in Table~\ref{results}
or \ref{maxmin}.
The implied rates are around 5\e{-2}~\solmass~\peryr, much higher than
the infall rates expected for low-mass star formation.
However, the infall rate onto the star itself
might be lower than the rate we estimate at these 0.1 pc scales.
Spin-up motions and stellar winds/outflows are both observed in e2 at
radii $<$ 0.01 pc (Zhang \& Ho 1997; Gaume et al. 1993).
Magnetic fields are undoubtedly also important.
There is an absence of good theoretical models of high-mass star
formation which would place our inferred mass infall rates in an
appropriate context.

\subsection{Observed features which are not reproduced}
\label{notreproduced}

All of the models underestimate the
strength of the main hyperfine component in emission, though they fit
the strength of the satellite components fairly well.  This discrepancy
seems to indicate that all of the models are lacking some hot, optically thin
gas.
In fact, Ho et al. (1983) measured molecular gas temperatures of
\approx\ 100~K in the core e2 using the (3,3) line of \ammonia. 
Our models, however, do not contain gas at such high temperatures.

The models also fail to reproduce some emission near
the center of the cloud, seen in Figure~\ref{comparetothis} at 
19\thr 21\tmin 26.25\tsec\ and
62 \kms, at a level of 36 mJy/beam  or 12~K (6$\sigma$).
The velocity of this gas is more redshifted than most of the gas seen in
absorption.  
Since this gas at 62 \kms\ is seen in emission, and our spatial resolution 
is much larger
than the actual size of the HII region, it is not possible to know if
the gas is in front of or behind the HII region.
This emission could come from gas behind the HII region;
in that case, its redshifted velocity suggests expansion
or outflow from the molecular core.
Thus, it is possible that the e2 molecular core is experiencing
simultaneous infall and outflow.

This emission at 62 \kms\ could also be explained by the presence of some hot,
optically thin gas in front of the HII region.
The brightness temperature of the HII region is about 80~K.
(The HII region is not resolved by the current observations; see Paper~I.)
The optical depths of the molecular gas in e2 are very high, 5--10. 
Thus, molecular gas in front of the HII region would need an excitation
temperature of only about 90~K in order to be seen in emission at 12~K
in front of the HII region.
As discussed above, Ho et al. (1983) did indeed find evidence of
temperatures around 100~K in the e2 molecular core.
Although the models of this paper do not favor an inside-out collapse
structure (see Section~\ref{quantitative}), the gas described here---
warmer gas, presumably closer to the HII region, and moving at higher
velocities--- may provide some evidence in favor of an inside-out collapse.
In any case, whether the gas at 62 \kms\ is infalling or outflowing,
it does not contradict the conclusion of Paper~I that the bulk of
the gas in e2 must be infalling.

Finally, none of the spherically symmetric models reproduces the
asymmetries apparent in the data.
Model 5b, with an offset HII region, is asymmetric but the
overall fit to the data (reduced \ch2) is not improved
(Section~\ref{model5section}).

\subsection{Uncertainties} \label{uncertainties}

The simple error analysis described in Section~\ref{procedure} gives
estimates of the $1\sigma$ uncertainties in the model parameters,
assuming the parameters are not correlated.  
Typical values for these uncertainties 
are presented in the last column of Table~\ref{results}.  
The errors of fitting are
typically quite small, and they do not reflect the true uncertainties
because they ignore systematic errors.
Major sources of error in this technique are the flux calibration of the
data, the distance uncertainty, the assumed \ammonia\ abundance,
and the interdependence of temperature and
density.

One source of systematic error is the 
uncertainty in the flux calibration of the VLA data and the 
primary beam correction.  
Changes in the flux calibration scale the brightness temperature 
by some multiplicative factor.
This scaling factor should translate into an uncertainty in the
temperature of the cloud, since the absolute
magnitude or strength of the lines should be determined 
largely by the temperature of the gas.  
Experience indicates that the uncertainty in the flux calibration of VLA
data may be as large as 20\% at K-band (23 GHz).
In addition, the primary beam correction could be as large as 30\% at the 
position of e2, though random pointing errors would tend to decrease the 
primary beam correction.

Another source of systematic error is the distance
uncertainty.  As mentioned earlier, the method of statistical parallax
applied to the masers in W51 gives a distance of $7.0 \pm 1.5$ kpc
(Genzel et al. 1982).  
This 21\% 
uncertainty in the distance to the cloud produces a 21\% uncertainty in the
linear radius of the cloud and hence the gas density $n_0$ (in order 
to produce the same total column density). 

Because the densities quoted here are molecular hydrogen densities, scaled
up from the data by an assumed \ammonia\ abundance, the unknown \ammonia\
abundance of course contributes to uncertainties in density.
We have adopted $\rm NH_3/H_2$ = 1.4\e{-6}, 
but this value is probably uncertain by at least a factor of 10 
(Ho \& Townes 1983).
If we had adopted an abundance value a factor of 10 smaller, the densities 
in Tables~\ref{parameters}, \ref{results}, and \ref{maxmin} would increase by that factor.
Moreover, the \ammonia\ abundance could vary with radius in the core.
An abundance gradient would mimic the effect of a density gradient, and the
present modelling technique cannot distinguish between the two.
We have also assumed that the \ammonia\ level populations are determined by
LTE.
If this assumption does not hold, the model gas densities and temperatures
would be inaccurate; however,
because of the complex source geometry, it is difficult to predict whether they
would be underestimated or overestimated.
At the high densities found in the e2 core, the LTE assumption is
likely to cause smaller uncertainties than those introduced by the \ammonia\ abundance.

In this modelling technique, it is difficult to make a unique determination
of kinetic temperature and volume density because of an inverse correlation
between these two quantities
(see also Section~\ref{model4section}).
This correlation arises from the fact that a spectrum of a single \nh3\ 
inversion transition constrains the optical depth of the transition
and the beam-diluted brightness temperature, not the volume density or
kinetic (or excitation) temperature (Ho \& Townes 1983).
The comparison between models 4a and 4b and between models 6a and 6b 
(Table~\ref{results}) shows that 
one can trade off a factor of two increase in temperature 
for a factor of 2 to 4 decrease in density and achieve the same \ch2.
Because good fits are not obtained for variations 
much beyond this range, we conclude that this interdependence
between temperature and density brackets the temperature to a factor
of two and the density to better than an order of magnitude.
Simultaneous fitting of more than one transition would remove much of the
ambiguity.

An inaccuracy of a few percent results from the coarse gridding of
the data in Figure~\ref{comparetothis}.
That is, the value of \ch2\ depends on 
exactly how the original data cube is sampled
because the final convolution of the model to approximate the
resolution of the VLA only uses 49
discrete points that cover an observing beam.
We find errors on the order of 6\% based on sampling.
In addition, 
uncertainties in the continuum level translate into
uncertainties in the temperature of the gas and continuum opacity of the HII
region via the strength of the absorption.
The
errors in continuum subtraction are probably on the order of 6 mJy/beam (2~K) or
less, as that is the rms noise in the line-free regions of the data.
In comparison, the strongest absorption in the core e2 is 120 mJy/beam.
Continuum subtraction probably produces relatively small errors.

The fitting procedure employed in this work is a local minimization
procedure, rather than a global minimization.
However, in practice a large amount of global minimization has already been
done.
The reason for this is that the output model is extremely sensitive to the
initial conditions, so that if the initial guess is not relatively good the
program tends to run the model down to blank sky instead of to a meaningful
fit.
Thus the various sets of initial conditions presented in Table~\ref{parameters}
are only a small subset of the ones that were attempted,
most of which gave unacceptable results.

\section{Conclusions}

We present radiative transfer modelling of the \nh3\ (J,K) = (2,2)
transition in the molecular core around the ultracompact HII region W51~e2.
Paper~I described the \nh3\ observations and presented a model in which the
molecular core (radius \approx 0.1 pc) is undergoing roughly
spherically symmetric contraction at about 5 \kms\ onto the young massive
star.
This paper investigates the physical properties and three-dimensional
structure of the core in more detail through numerical techniques,
using a series of models of gradually increasing complexity in
which the gas temperature, density, and infall velocity are parametrized as
power laws.
The parameters of these models were optimized so that the expected
line radiation best matched the observed data.

Comparison of the series of models yields insights into the importance of
various model parameters.
For example, the core is contracting at a velocity of about 5 \kms.
A good  model of the core requires that the temperature and density of
the gas both decrease with increasing distance from the center of the cloud,
over  0.1~pc scales.
Major uncertainties arise from the assumed
\ammonia\ abundance and from the fact that the temperature and density
cannot be determined independently in this project.
The flux calibration of the data and the distance to W51 also
introduce significant uncertainties.
An important feature of this work is that, regardless of the numerical
uncertainties, comparing models of gradually
increasing complexity yields insights into 
the sensitivity of the model to the parameters and indicates which
parameters are most important.
For example,
models without infall and
isothermal models are clearly inadequate descriptions of
the molecular core.

\acknowledgments

We thank the anonymous referee for a thorough discussion of the paper.
Work on this paper started when LMY was an undergraduate at Harvard
University.

\newpage

\begin{deluxetable}{lrrrrrrrrr}  
\footnotesize
\tablewidth{0pt}
\tablecaption{Initial guesses for models \label{parameters}}

\tablehead{
\colhead{Property}  &  \multicolumn{9}{c}{Model}  \nl
\colhead{} & \colhead{1} & \colhead{2} & \colhead{3} & \colhead{4a} &
\colhead{4b} & \colhead{5a} & \colhead{5b} & \colhead{6a} & \colhead{6b} 
}

\startdata
$\rm R_{inner}$ (pc)&
 0.010 & 0.030 & 0.030 & 0.030 & 0.030 & 0.030 & 0.030 & 0.030 & 0.030 \nl
$\rm R_{outer}$ (pc)&
 0.18 & 0.18 & 0.18 & 0.18 & 0.18 & 0.18 & 0.18 & 0.18 & 0.18 \nl
$T_0$ (K)&
 12.6 & 12.6 & 8.2 & 12.6 & 25.0 & 12.6 & 20.0 & 12.6 & 25.0 \nl
$\alpha_T$&
 \nodata & \nodata & \nodata & $-0.5$ & $-0.5$ & $-0.5$ & $-0.6$ &
 $-0.5$ & $-0.5$ \nl
$n_0$ (\tenup{6} \percc)&
 1.1 & 3.0 & 59 & 3.0 & 1.0 & 11 & 1.1 & 3.0 & 1.0 \nl
$\alpha_n$&
 \nodata & \nodata & $-1.5$\tablenotemark{a} & $-2.0$ & $-2.0$ & $-1.7$ &
 $-2.0$ & $-2.0$ & $-2.0$ \nl
$v_0$ (\kms)&
 \nodata & 5.0 & 4.5 & 5.0 & 5.0 & 5.0 & 5.0 & 5.0 & 5.0 \nl
$\alpha_v$&
 \nodata & $-0.5$ & $-0.5$\tablenotemark{a} & $-0.5$ & 0.0 & $-0.5$ & 0.0 &
 0.0 & 0.0 \nl
$\rm offset_1$ (pc)&
 \nodata & \nodata & \nodata & \nodata & \nodata & 0.0 & 0.05 &
 \nodata & \nodata \nl
$\rm offset_2$ (pc)&
 \nodata & \nodata & \nodata & \nodata & \nodata & 0.0 & 0.05 &
 \nodata & \nodata \nl
aspect ratio&
 \nodata & \nodata & \nodata & \nodata & \nodata & \nodata & \nodata & 
 1.0 & 4.0 \nl
inclination (\deg)&
 \nodata & \nodata & \nodata & \nodata & \nodata & \nodata & \nodata & 
 0 & 45 \nl
\enddata

\tablenotetext{a}{fixed}
\tablecomments{Starting parameter values are presented for each model.
Two free parameters are not shown in this table:  the continuum opacity of
the HII region, and the velocity of the ionized gas in the HII region.
Parameters $\rm R_{inner}$ and $\rm R_{outer}$ refer to the inner and outer
edges of the shell of molecular gas.
Parameters $T_0$,  $n_0$ and $v_0$ are temperature, density (molecular
hydrogen), and infall
velocity at radius $r_0$ = 0.05 pc from the center of the HII region, except
in models 5a and 5b (see text).
Offset$_1$ is a displacement of the HII region along the line of sight, and
offset$_2$ is along the direction of right ascension.
}

\end{deluxetable}

\begin{deluxetable}{lrrrrrrrrrrrrr} 
\footnotesize
\tablewidth{0pt}
\tablecaption{Optimized parameters of the models \label {results}}

\tablehead{
\colhead{Property}  &  \multicolumn{9}{c}{Model}  &  \colhead{Error} \nl
\colhead{} & \colhead{1} & \colhead{2} & \colhead{3} & \colhead{4a} &
\colhead{4b} & \colhead{5a} & \colhead{5b} & \colhead{6a} & \colhead{6b} &
\colhead{}
}

\startdata
$\rm R_{inner}$ (pc)&
 0.003 & 0.027 & 0.027 & 0.028 & 0.024 & 0.030 & 0.028 & 0.029 & 0.023 & 
 0.002 \nl
$\rm R_{outer}$ (pc)&
 0.12 & 0.09 & 0.15 & 0.18 & 0.12 & 0.18 & 0.11 & 0.21 & 0.18 & 0.03 \nl
$T_0$ (K)&
 12.5 & 8.2 & 8.2 & 12.8 & 17.9 & 12.4 & 12.3 & 13.0 & 24.3 & 0.5 \nl
$\alpha_T$&
 \nodata & \nodata & \nodata & $-0.61$ & $-0.51$ & $-0.60$ & $-0.61$ &
 $-0.58$ & $-0.64$ & 0.05 \nl
$n_0$ (\tenup{6} \percc)&
 0.31 & 59 & 23 & 6.8 & 1.5 & 12.3 & 22 & 6.9 & 2.4 & 10\% \nl
$\alpha_n$&
 \nodata & \nodata & $-1.5$\tablenotemark{a} & $-1.8$ & $-3.9$ & $-2.0$ &
 +0.9 & $-2.3$ & $-2.2$ & 0.2 \nl
$v_0$ (\kms)&
 \nodata & 4.5 & 4.3 & 4.8 & 4.7 & 4.8 & 6.6 & 4.9 & 4.0 & 0.05 \nl
$\alpha_v$&
 \nodata & 0.12 & $-0.5$\tablenotemark{a} & 0.09 & $-0.02$ & 0.01 & 0.18 &
 0.10 & 0.10 & 0.03 \nl
$\rm offset_1$ (pc)&
 \nodata & \nodata & \nodata & \nodata & \nodata & 0.001 & 0.051 &
 \nodata & \nodata & 0.005 \nl
$\rm offset_2$ (pc)&
 \nodata & \nodata & \nodata & \nodata & \nodata & 0.005 & 0.051 &
 \nodata & \nodata & 0.005\nl
aspect ratio&
 \nodata & \nodata & \nodata & \nodata & \nodata & \nodata & \nodata & 
 1.0 & 10.3 & 0.5 \nl
inclination (\deg)&
 \nodata & \nodata & \nodata & \nodata & \nodata & \nodata & \nodata & 
 15 & 0 & 5 \nl
\ch2 &
 11.2 & 3.8 & 4.1 & 3.2 & 3.0 & 3.1 & 2.7 & 3.2 & 3.2 & \nl
\enddata

\tablenotetext{a}{fixed}

\end{deluxetable}

\begin{deluxetable}{llllllllll} 
\footnotesize
\tablewidth{0pt}
\tablecaption{Physical properties of the molecular gas \label{maxmin}}

\tablehead{
\colhead{Property} & \multicolumn{9}{c}{Model}    \nl
\colhead{} & \colhead{1} & \colhead{2} & \colhead{3} & \colhead{4a} &
\colhead{4b} & \colhead{5a} & \colhead{5b} & \colhead{6a} & \colhead{6b} 
}

\startdata
inner radius (pc)&  
0.003 & 0.027 & 0.027 & 0.028 & 0.024 & 0.030 & 0.028 & 0.029 & 0.023\nl
outer radius (pc)&  
0.12&   0.089&   0.15 & 0.18&  0.12& 0.18&   0.11& 0.21&  0.18 \nl
$T$, max. (K)& 
13 &   8.2&  8.2 & 18& 26&  17&  18&  18& 40\nl
$T$, min. (K)& 
13&   8.2&  8.2 &  5.9& 12&  5.7&   5.6&   5.7& 11\nl
$n$, max. (\tenup{6} \percc)&  
0.31&  59&  57 & 20 & 25 & 35&  46&  25& 13\nl
$n$, min. (\tenup{6} \percc)&  
0.31&  59&  4.2 &  0.68&  0.049&  0.95&   0.074&   0.25&  0.15\nl
$v$, max. (km/s)&  
0&   4.8&   5.9 & 5.4&  4.8&  4.8&   7.6&   5.6&  6.0\nl
$v$, min. (km/s)&  
0&   4.1&   2.5 & 4.6&  4.6&   4.7&   2.1&   4.6&  3.7\nl
Gas mass (\solmass)&
120& 8500& 6300& 1800& 190& 2900& 9300& 1600& 540\nl
\enddata

\tablecomments{Physical properties of the gas in each of the 
models.}
\end{deluxetable}

\begin{deluxetable}{llllllllll}  
\tablewidth{0pt}
\tablecaption{Energy balance calculations \label{energybal}}

\tablehead{
\colhead{Energy} & \multicolumn{9}{c}{Model}   \nl
\colhead{(\tenup{48} erg)} & \colhead{1} & \colhead{2} & \colhead{3} & \colhead{4a} &
\colhead{4b} & \colhead{5a} & \colhead{5b} & \colhead{6a} & \colhead{6b} 
}

\startdata
Potential         &
0.01  &40   & 10    &1   &0.04  &3   &40       &0.9 &0.1  \nl
Kinetic           &
0     &\phd 2   & \phd 0.7 & 0.5 &0.04 &0.7 &\phd 5   &0.4 &0.1  \nl
Thermal/turbulent &
0.008 &\phd 0.6 & \phd 0.4 & 0.1 &0.01 &0.2 &\phd 0.6 &0.1 &0.04 \nl
\enddata

\tablecomments{Total 
energy in the form of gravitational potential, kinetic, and thermal energy 
integrated over the entire molecular core (see text).}

\end{deluxetable}

\end{document}